\documentclass[aps,prl,twocolumn,showpacs,amssymb]{revtex4-1}
\usepackage{graphicx}
\usepackage{tabularx}

\begin{document}

\title{Mechanical cleaning of graphene}

\author{A.M. Goossens$^1$, V.E. Calado$^1$, A. Barreiro$^1$, K. Watanabe$^2$, T. Taniguchi$^2$, L.M.K. Vandersypen$^1$}
\email{Corresponding author, l.m.k.vandersypen@tudelft.nl}
\affiliation{$^1$Kavli Institute of Nanoscience, Delft University of Technology, P.O. Box 5046, 2600 GA Delft, The Netherlands, $^2$Advanced Materials Laboratory, National Institute for Materials Science, 1-1 Namiki, Tsukuba, 305-0044, Japan}
\date{\today}

\begin{abstract}
Contamination of graphene due to residues from nanofabrication often introduces background doping and reduces electron mobility. For samples of high electronic quality, post-lithography cleaning treatments are therefore needed. We report that mechanical cleaning based on contact mode AFM removes residues and significantly improves the electronic properties. A mechanically cleaned dual-gated bilayer graphene transistor with hBN dielectrics exhibited a mobility of $\sim$36,000 cm$^2$/Vs at low temperature. 
\end{abstract}

\maketitle

High electronic quality is demanded for many graphene experiments~\cite{neto_electronic_2009,fuhrer_graphene:_2010}, but is not easily realized. Graphene samples for electronic measurements are typically made with lithographic methods. Lithography makes a myriad of devices possible, but always leaves resist residues behind. Making contacts to graphene with shadow mask evaporation solves this contamination issue, but this method has many drawbacks concerning the flexibility of the fabrication process. Hence cleaning after lithography is a crucial step towards obtaining high electronic quality samples. There are different methods at hand: chemical cleaning ~\cite{cheng_toward_2011}, thermal cleaning (annealing in an oven) ~\cite{ishigami_atomic_2007,dan_intrinsic_2009} and current-induced cleaning ~\cite{moser_current-induced_2007,bolotin_ultrahigh_2008}. Each of these can be very useful but has its own limitations. 

In this paper we present an alternative cleaning method: mechanical cleaning. Scanning a contact mode AFM (CM AFM) tip over a graphene surface removes residues, removes doping and improves the electronic mobility without damaging the graphene, 

\begin{figure}[b]
\begin{center}
\includegraphics[width=8.5cm]{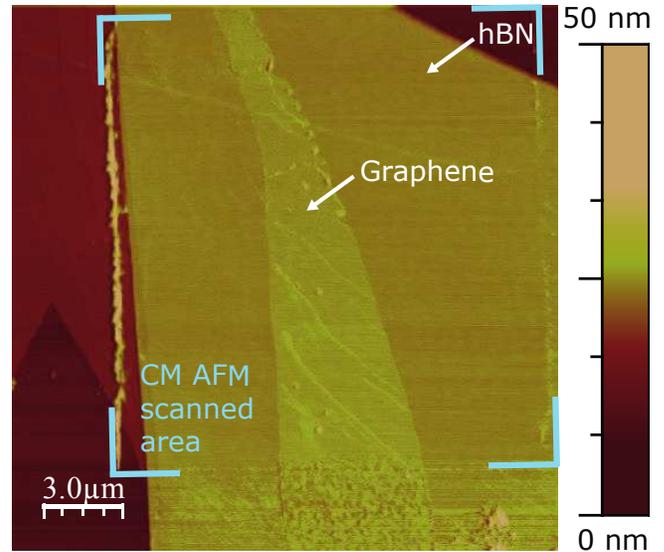}
\end{center}
\vspace*{-.5cm}
\caption{(color online) Tapping mode image of sample A after annealing at $440~^{\circ}$C and contact mode scanning (both with a Veeco Nanoscope IIIa AFM). Only the part within the marked window was scanned with the CM AFM. We chose to show this device because it was much more contaminated than other devices before scanning, so that the effect of the CM AFM scan is easily visible. Wrinkles and some tears on the upper right side of the graphene are induced by the tip but were not observed in other devices. On the left and right of the bounding box, walls of deposited residue are visible. The contacts of the device are not visible in this image.}
\vspace*{-.3cm}
\label{fig:afm}
\end{figure}

We demonstrate the effectiveness of this method for 4 bilayer graphene on hexagonal boron nitride (hBN) samples. hBN flakes are deposited by mechanical exfoliation on silicon wafers coated with a silicon oxide (SiO$_{2}$) layer of thickness $t_{SiO_{2}}=285$ nm. On top of the hBN we transfer a bilayer graphene flake using a dry transfer method following the protocol of ~\cite{dean_boron_2010} (at a temperature of $100~^{\circ}$C to remove any water absorbed on the surface of the graphene and hBN flakes). Samples are subsequently annealed in an oven at $400~^{\circ}$C (Ar $2400$ sccm, H$_{2}$ $700$ sccm) to remove residues induced by the transfer process. Cr/Au electrodes are fabricated using electron-beam lithography. We annealed the samples again (same flow rate as the first annealing step) to remove fabrication residues. Trying to clean the graphene, we performed multiple annealing steps at temperatures from $300~^{\circ}$C to a maximum of $440~^{\circ}$C. 

After the final annealing step, the samples were often still contaminated. The tapping mode AFM (TM AFM) image of sample A (Fig. ~\ref{fig:afm}) shows lots of deposited material outside the marked window. The roughness in this area is $\sim1$ nm. Before lithography all samples were almost atomically flat with a roughness of at most $0.2$ nm (limited by the resolution of the AFM).

\begingroup
\squeezetable
\begin{table*}[t!]	
\renewcommand{\arraystretch}{1.5}	
\centering		
\begin{tabularx}{\textwidth}{XXXXXXXXX}		
Sample & Anneal T & Nr of passes & Scan force  & Measurement & $V_{np}$ before & $V_{np}$ after & $\mu$ before & $\mu$ after  \\	
 &[$^\circ$C] &  & [nN] & & [V] & [V] & [cm$^2$/Vs] & [cm$^2$/Vs] \\
\hline		
A & 440 & 6 & 2.3 & 2-prb & 4 & -7 & $3.4\cdot10^{3}$ & $8.9\cdot10^{3}$   \\	
B & 360 & 1 & -2.9 & 3-prb & $>$20 & 0 & $1.7\cdot10^3$ & $2.8\cdot10^{3}$    \\	
C & 360 & 2 & -4.6 & 3-prb  & $>$20 & 1 & $2.6\cdot10^2$ & $9.2\cdot10^{2}$   \\	
D & 360 & 5 & -22 & 4-prb & 17 & -1 & $2.7\cdot10^{3}$ & $6.7\cdot10^{3}$   \\								
\end{tabularx}	
\caption{Results summary for four different samples (measurements at room temperature in vacuum).}
\label{tbl:table1}
\end{table*}
\endgroup

\begin{figure}[b]
\begin{center}
\includegraphics[width=7cm]{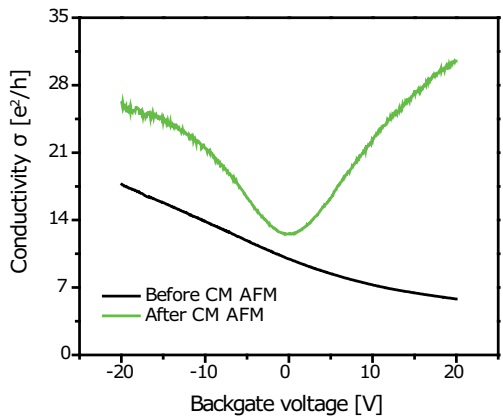}
\end{center}
\vspace*{-.7cm}
\caption{ (color online) \textbf{a)} Backgate traces of sample B at room temperature in vacuum ($I_{bias}=100$ nA). The black curve is before CM AFM imaging and the green curve after. Using the geometric capacitance, we calculated the carrier density from this plot and then extracted the field effect mobility by fitting a straight line to the steepest part of the backgate trace: $\mu=(\frac{t_{SiO_{2}}}{\epsilon_{0}\epsilon_{r,SiO_{2}}}+\frac{t_{hBN}}{\epsilon_{0}\epsilon_{r,hBN}})\frac{d\sigma}{dV}$, where $\epsilon_{r,SiO_{2}}=3.9$ and $\epsilon_{r,hBN}=3.0$, as calculated from Fig.~\ref{fig:2dplot}. We extracted the thickness t$_{hBN}$ from TM AFM images of the devices.}
\vspace*{-.3cm}
\label{fig:backgate}
\end{figure}

We characterize the electronic quality of the samples via the residual doping and field effect mobility. Fig.~\ref{fig:backgate} (black trace) shows the resistance of sample B as a function of the backgate voltage at room temperature in vacuum. From this and similar traces for the other devices, we extracted the charge neutrality point ($V_{np}$) and mobility ($\mu$) of the samples. Depending on the device we were able to do 2,3 or 4 terminal measurements. In the two and three terminal measurements, contact resistances make the mobility appear lower. For the aspect ratio of the flake, we always took an underestimate. Those two factors make the measured mobility a lower bound of the actual mobility. The results for all samples are summarized in table~\ref{tbl:table1}. The electronic measurements indicate that most of the devices were highly doped and had a relatively low mobility. Most likely residues on top of the graphene as seen in the AFM images induce doping and provide scattering centers that degrade the electronic quality.~\cite{chen_charged-impurity_2008}

To remove the residues we scanned the samples in contact mode AFM with a constant force (Veeco OTR8-35 tip with a stiffness of $0.15$ N/m). Hereby the tip is held in contact with the sample surface. We engaged the tip with the lowest force possible. When the tip made contact, we confirmed a reasonable set-point force with the help of a force distance measurement, discussed further below. Then we started scanning the sample with a rate of $0.5-1$ Hz. For most samples we scanned the same area several times, but without further visible effect.

Tapping mode images taken after scanning in CM AFM show that we cleaned the graphene (Fig.~\ref{fig:afm} inside marked window). The roughness is at most $0.2$ nm, similar to the values measured before processing the devices. Further evidence that we removed residue from the graphene are the banks of deposits that are visible in Fig.~\ref{fig:afm}, exactly at the boundaries of the area that was scanned in contact mode. 

After CM and TM AFM imaging we again recorded backgate traces at room temperature in vacuum (green curve in Fig.~\ref{fig:backgate}). Not only the mobility increased twofold, but also doping was reduced. For other samples we observed similar behavior (see table~\ref{tbl:table1}). 

\begin{figure}[b!]
\begin{center}
\includegraphics[width=8.5cm]{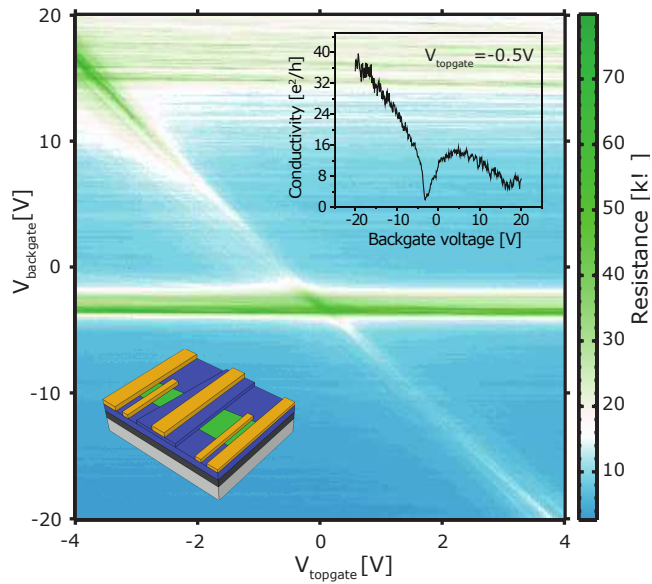}
\end{center}
\vspace*{-.5cm}
\caption{(color online) Measurements on a double gated bilayer graphene transistor fabricated out of sample A. The 4-probe resistance at T = 50 mK is plotted as a function of backgate and topgate voltage. From the slope of the diagonal line we calculated the relative dielectric constant of the hBN to be $3.0$ assuming a parallel plate capacitor model and $\epsilon_{r,SiO_{2}}=3.9$. The thickness of the bottom hBN flake was $14$ nm and the top hBN flake $50$ nm, values extracted from AFM images. Lowerleft inset: schematic of the device. Blue colored regions are hBN, green is bilayer graphene and yellow are the contacts and gate. Upperright inset: resistance as a function of backgate voltage at $V_{topgate}=-0.5$V. The dip at $V_{bg}\sim17$V is caused by the uncovered graphene part.}
\vspace*{-.3cm}
\label{fig:2dplot}
\end{figure}

An attractive feature of mechanical cleaning is that it can be naturally followed by further sample processing. We fabricated sample A into a double gated bilayer device~\cite{oostinga_gate-induced_2008, zhang_direct_2009}. With the same dry transfer method as mentioned before we stamped a hBN flake on sample A that will act as a topgate dielectric. We defined a topgate electrode across the flake and two voltage probes by e-beam lithography (lower left inset Fig.~\ref{fig:2dplot}). Resistance as function of the topgate and backgate voltages is plotted in Fig.~\ref{fig:2dplot}. From the upper right inset we extract a (hole) mobility $\mu$ of $\sim 36,000$ cm$^2$/Vs at carrier density n$\sim5*10^{10}$ cm$^{-2}$ (corrected for the change in slope due to the neutrality point around $17$ V). This value is among the highest found in the literature for bilayer graphene devices, including suspended devices ~\cite{dean_boron_2010,feldman_broken-symmetry_2009, mayorov_interaction-driven_2011}. In the combined topgate and backgate sweep data the diagonal shows the typical increase of resistance with increasing perpendicular electric field~\cite{oostinga_gate-induced_2008}.

We now turn to the mechanism by which CM AFM removes residues from the sample surface. Presumably the tip is plowing through a layer of physisorbed contaminants and thereby 'brooms' the graphene clean, which means the interaction of the tip with the surface is larger than the interaction of the contaminants with the surface. We believe that removing residues is the main explanation for the improvement of the electronic quality of the graphene. In principle the CM AFM might also flatten the graphene, reducing ripples and thereby enhancing mobility. However, we observe no difference in the flatness of graphene before fabrication (presumably equal to the state after lithography) and after mechanical cleaning. Flattening of the graphene should thus play little or no role. The hBN substrate does appear to play a role in improving electronic quality. Jalilian et al.~\cite{jalilian_scanning_2011} deployed the mechanical cleaning on single layer graphene samples on a SiO$_{2}$ substrate. They also observed an improvement in surface morphology, but electronic quality did not improve. We observed the same behavior in a single layer graphene on SiO$_{2}$ sample. Further research needs to be done to explain the role of the substrate.

\begin{figure}[t!]
\begin{center}
\includegraphics[width=8.5cm]{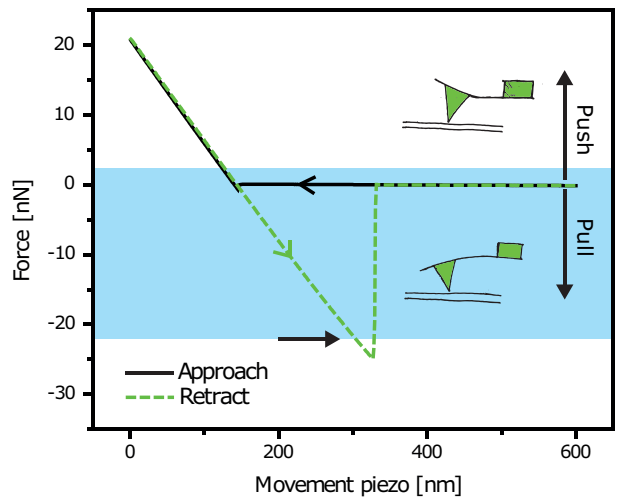}
\end{center}
\vspace*{-.5cm}
\caption{(color online) Force-distance curve of sample D, measured by holding the tip of the AFM in a fixed lateral position and approaching and retracting the tip in the vertical direction. While making these vertical movements, the deflection of the tip is recorded. Assuming that when the tip is in contact with the surface the tip deflects the same distance as the piezo moves, we can calibrate the deflection scale. With the spring constant of the tip we convert that deflection to a force. The horizontal axis has an arbitrary offset. The blue region indicates the range of forces that we used for cleaning the samples. Sample D was scanned at a force of $-22$ nN as indicated by the arrow. The illustrations picture the pulling and pushing regime.}
\vspace*{-.3cm}
\label{fig:fdcurve}
\end{figure}

To gain more insight in the interplay of the surface and the CM AFM tip we took force-distance curves (Fig.~\ref{fig:fdcurve}) in the area we scanned in contact mode. From these curves we can extract the force we were exerting on the sample during scanning, which ranged from $-22$ nN to $+2.3$ nN depending on the device. A positive force means the tip was pushing and a negative force that the tip was pulling on the surface. In pulling configuration the tip is held in contact by the Van der Waal's interaction and adhesive forces due to water. The broad range of scanning forces that gave good results illustrates the robustness of the mechanical cleaning method.

In summary, scanning bilayer graphene on hBN in CM AFM removes contaminants from the surface, reduces residual doping and significantly improves electronic mobility. A double gated bilayer graphene transistor which was mechanically cleaned in the fabrication process, showed mobilities up to $36,000$ cm$^{2}$/Vs at $50$ mK, and opening of a bandgap. This illustrates the effectiveness and versatility of CM AFM for obtaining high-quality graphene devices. Possibly, AFM and STM setups in vacuum could benefit even more from mechanical cleaning as it can be applied in situ, avoiding subsequent contamination by molecules absorbed from the air ~\cite{ishigami_atomic_2007,castellanos-gomez_carbon_2010}. 

We acknowledge useful discussions with A. Castellanos-Gomez and T. van der Sar, experimental support from L. Cantley and financial support from the Foundation for Fundamental Research on Matter (FOM) and the European Research Council (ERC).

\bibliography{grafeengum}

\begin{thebibliography}{15}%
\makeatletter
\providecommand \@ifxundefined [1]{%
 \@ifx{#1\undefined}
}%
\providecommand \@ifnum [1]{%
 \ifnum #1\expandafter \@firstoftwo
 \else \expandafter \@secondoftwo
 \fi
}%
\providecommand \@ifx [1]{%
 \ifx #1\expandafter \@firstoftwo
 \else \expandafter \@secondoftwo
 \fi
}%
\providecommand \natexlab [1]{#1}%
\providecommand \enquote  [1]{``#1''}%
\providecommand \bibnamefont  [1]{#1}%
\providecommand \bibfnamefont [1]{#1}%
\providecommand \citenamefont [1]{#1}%
\providecommand \href@noop [0]{\@secondoftwo}%
\providecommand \href [0]{\begingroup \@sanitize@url \@href}%
\providecommand \@href[1]{\@@startlink{#1}\@@href}%
\providecommand \@@href[1]{\endgroup#1\@@endlink}%
\providecommand \@sanitize@url [0]{\catcode `\\12\catcode `\$12\catcode
  `\&12\catcode `\#12\catcode `\^12\catcode `\_12\catcode `\%12\relax}%
\providecommand \@@startlink[1]{}%
\providecommand \@@endlink[0]{}%
\providecommand \url  [0]{\begingroup\@sanitize@url \@url }%
\providecommand \@url [1]{\endgroup\@href {#1}{\urlprefix }}%
\providecommand \urlprefix  [0]{URL }%
\providecommand \Eprint [0]{\href }%
\providecommand \doibase [0]{http://dx.doi.org/}%
\providecommand \selectlanguage [0]{\@gobble}%
\providecommand \bibinfo  [0]{\@secondoftwo}%
\providecommand \bibfield  [0]{\@secondoftwo}%
\providecommand \translation [1]{[#1]}%
\providecommand \BibitemOpen [0]{}%
\providecommand \bibitemStop [0]{}%
\providecommand \bibitemNoStop [0]{.\EOS\space}%
\providecommand \EOS [0]{\spacefactor3000\relax}%
\providecommand \BibitemShut  [1]{\csname bibitem#1\endcsname}%
\let\auto@bib@innerbib\@empty
%</preamble>
\bibitem [{\citenamefont {Neto}\ \emph {et~al.}(2009)\citenamefont {Neto},
  \citenamefont {Guinea}, \citenamefont {Peres}, \citenamefont {Novoselov},\
  and\ \citenamefont {Geim}}]{neto_electronic_2009}%
  \BibitemOpen
  \bibfield  {author} {\bibinfo {author} {\bibfnamefont {A.~H.~C.}\
  \bibnamefont {Neto}}, \bibinfo {author} {\bibfnamefont {F.}~\bibnamefont
  {Guinea}}, \bibinfo {author} {\bibfnamefont {N.~M.~R.}\ \bibnamefont
  {Peres}}, \bibinfo {author} {\bibfnamefont {K.~S.}\ \bibnamefont
  {Novoselov}}, \ and\ \bibinfo {author} {\bibfnamefont {A.~K.}\ \bibnamefont
  {Geim}},\ }\href {\doibase 10.1103/RevModPhys.81.109} {\bibfield  {journal}
  {\bibinfo  {journal} {Reviews of Modern Physics}\ }\textbf {\bibinfo {volume}
  {81}},\ \bibinfo {pages} {109} (\bibinfo {year} {2009})}\BibitemShut
  {NoStop}%
\bibitem [{\citenamefont {Fuhrer}\ \emph {et~al.}(2010)\citenamefont {Fuhrer},
  \citenamefont {Lau},\ and\ \citenamefont
  {{MacDonald}}}]{fuhrer_graphene:_2010}%
  \BibitemOpen
  \bibfield  {author} {\bibinfo {author} {\bibfnamefont {M.~S.}\ \bibnamefont
  {Fuhrer}}, \bibinfo {author} {\bibfnamefont {C.~N.}\ \bibnamefont {Lau}}, \
  and\ \bibinfo {author} {\bibfnamefont {A.~H.}\ \bibnamefont {{MacDonald}}},\
  }\href {\doibase 10.1557/mrs2010.551} {\bibfield  {journal} {\bibinfo
  {journal} {{MRS} Bulletin}\ }\textbf {\bibinfo {volume} {35}},\ \bibinfo
  {pages} {289} (\bibinfo {year} {2010})}\BibitemShut {NoStop}%
\bibitem [{\citenamefont {Cheng}\ \emph {et~al.}(2011)\citenamefont {Cheng},
  \citenamefont {Zhou}, \citenamefont {Wang}, \citenamefont {Li}, \citenamefont
  {Wang},\ and\ \citenamefont {Fang}}]{cheng_toward_2011}%
  \BibitemOpen
  \bibfield  {author} {\bibinfo {author} {\bibfnamefont {Z.}~\bibnamefont
  {Cheng}}, \bibinfo {author} {\bibfnamefont {Q.}~\bibnamefont {Zhou}},
  \bibinfo {author} {\bibfnamefont {C.}~\bibnamefont {Wang}}, \bibinfo {author}
  {\bibfnamefont {Q.}~\bibnamefont {Li}}, \bibinfo {author} {\bibfnamefont
  {C.}~\bibnamefont {Wang}}, \ and\ \bibinfo {author} {\bibfnamefont
  {Y.}~\bibnamefont {Fang}},\ }\href {\doibase 10.1021/nl103977d} {\bibfield
  {journal} {\bibinfo  {journal} {Nano Lett.}\ }\textbf {\bibinfo {volume}
  {11}},\ \bibinfo {pages} {767} (\bibinfo {year} {2011})}\BibitemShut
  {NoStop}%
\bibitem [{\citenamefont {Ishigami}\ \emph {et~al.}(2007)\citenamefont
  {Ishigami}, \citenamefont {Chen}, \citenamefont {Cullen}, \citenamefont
  {Fuhrer},\ and\ \citenamefont {Williams}}]{ishigami_atomic_2007}%
  \BibitemOpen
  \bibfield  {author} {\bibinfo {author} {\bibfnamefont {M.}~\bibnamefont
  {Ishigami}}, \bibinfo {author} {\bibfnamefont {J.~H.}\ \bibnamefont {Chen}},
  \bibinfo {author} {\bibfnamefont {W.~G.}\ \bibnamefont {Cullen}}, \bibinfo
  {author} {\bibfnamefont {M.~S.}\ \bibnamefont {Fuhrer}}, \ and\ \bibinfo
  {author} {\bibfnamefont {E.~D.}\ \bibnamefont {Williams}},\ }\href {\doibase
  10.1021/nl070613a} {\bibfield  {journal} {\bibinfo  {journal} {Nano Lett.}\
  }\textbf {\bibinfo {volume} {7}},\ \bibinfo {pages} {1643} (\bibinfo {year}
  {2007})}\BibitemShut {NoStop}%
\bibitem [{\citenamefont {Dan}\ \emph {et~al.}(2009)\citenamefont {Dan},
  \citenamefont {Lu}, \citenamefont {Kybert}, \citenamefont {Luo},\ and\
  \citenamefont {Johnson}}]{dan_intrinsic_2009}%
  \BibitemOpen
  \bibfield  {author} {\bibinfo {author} {\bibfnamefont {Y.}~\bibnamefont
  {Dan}}, \bibinfo {author} {\bibfnamefont {Y.}~\bibnamefont {Lu}}, \bibinfo
  {author} {\bibfnamefont {N.~J.}\ \bibnamefont {Kybert}}, \bibinfo {author}
  {\bibfnamefont {Z.}~\bibnamefont {Luo}}, \ and\ \bibinfo {author}
  {\bibfnamefont {A.~T.~C.}\ \bibnamefont {Johnson}},\ }\href {\doibase
  10.1021/nl8033637} {\bibfield  {journal} {\bibinfo  {journal} {Nano Lett.}\
  }\textbf {\bibinfo {volume} {9}},\ \bibinfo {pages} {1472} (\bibinfo {year}
  {2009})}\BibitemShut {NoStop}%
\bibitem [{\citenamefont {Moser}\ \emph {et~al.}(2007)\citenamefont {Moser},
  \citenamefont {Barreiro},\ and\ \citenamefont
  {Bachtold}}]{moser_current-induced_2007}%
  \BibitemOpen
  \bibfield  {author} {\bibinfo {author} {\bibfnamefont {J.}~\bibnamefont
  {Moser}}, \bibinfo {author} {\bibfnamefont {A.}~\bibnamefont {Barreiro}}, \
  and\ \bibinfo {author} {\bibfnamefont {A.}~\bibnamefont {Bachtold}},\ }\href
  {\doibase 10.1063/1.2789673} {\bibfield  {journal} {\bibinfo  {journal}
  {Applied Physics Letters}\ }\textbf {\bibinfo {volume} {91}},\ \bibinfo
  {pages} {163513} (\bibinfo {year} {2007})}\BibitemShut {NoStop}%
\bibitem [{\citenamefont {Bolotin}\ \emph {et~al.}(2008)\citenamefont
  {Bolotin}, \citenamefont {Sikes}, \citenamefont {Jiang}, \citenamefont
  {Klima}, \citenamefont {Fudenberg}, \citenamefont {Hone}, \citenamefont
  {Kim},\ and\ \citenamefont {Stormer}}]{bolotin_ultrahigh_2008}%
  \BibitemOpen
  \bibfield  {author} {\bibinfo {author} {\bibfnamefont {K.}~\bibnamefont
  {Bolotin}}, \bibinfo {author} {\bibfnamefont {K.}~\bibnamefont {Sikes}},
  \bibinfo {author} {\bibfnamefont {Z.}~\bibnamefont {Jiang}}, \bibinfo
  {author} {\bibfnamefont {M.}~\bibnamefont {Klima}}, \bibinfo {author}
  {\bibfnamefont {G.}~\bibnamefont {Fudenberg}}, \bibinfo {author}
  {\bibfnamefont {J.}~\bibnamefont {Hone}}, \bibinfo {author} {\bibfnamefont
  {P.}~\bibnamefont {Kim}}, \ and\ \bibinfo {author} {\bibfnamefont
  {H.}~\bibnamefont {Stormer}},\ }\href {\doibase 10.1016/j.ssc.2008.02.024}
  {\bibfield  {journal} {\bibinfo  {journal} {Solid State Communications}\
  }\textbf {\bibinfo {volume} {146}},\ \bibinfo {pages} {351} (\bibinfo {year}
  {2008})}\BibitemShut {NoStop}%
\bibitem [{\citenamefont {Dean}\ \emph {et~al.}(2010)\citenamefont {Dean},
  \citenamefont {Young}, \citenamefont {Meric}, \citenamefont {Lee},
  \citenamefont {Wang}, \citenamefont {Sorgenfrei}, \citenamefont {Watanabe},
  \citenamefont {Taniguchi}, \citenamefont {Kim}, \citenamefont {Shepard},\
  and\ \citenamefont {Hone}}]{dean_boron_2010}%
  \BibitemOpen
  \bibfield  {author} {\bibinfo {author} {\bibfnamefont {C.}~\bibnamefont
  {Dean}}, \bibinfo {author} {\bibfnamefont {A.}~\bibnamefont {Young}},
  \bibinfo {author} {\bibfnamefont {I.}~\bibnamefont {Meric}}, \bibinfo
  {author} {\bibfnamefont {C.}~\bibnamefont {Lee}}, \bibinfo {author}
  {\bibfnamefont {L.}~\bibnamefont {Wang}}, \bibinfo {author} {\bibfnamefont
  {S.}~\bibnamefont {Sorgenfrei}}, \bibinfo {author} {\bibfnamefont
  {K.}~\bibnamefont {Watanabe}}, \bibinfo {author} {\bibfnamefont
  {T.}~\bibnamefont {Taniguchi}}, \bibinfo {author} {\bibfnamefont
  {P.}~\bibnamefont {Kim}}, \bibinfo {author} {\bibfnamefont {K.}~\bibnamefont
  {Shepard}}, \ and\ \bibinfo {author} {\bibfnamefont {J.}~\bibnamefont
  {Hone}},\ }\href {\doibase 10.1038/nnano.2010.172} {\bibfield  {journal}
  {\bibinfo  {journal} {Nat Nano}\ }\textbf {\bibinfo {volume} {5}},\ \bibinfo
  {pages} {722} (\bibinfo {year} {2010})}\BibitemShut {NoStop}%
\bibitem [{\citenamefont {Chen}\ \emph {et~al.}(2008)\citenamefont {Chen},
  \citenamefont {Jang}, \citenamefont {Adam}, \citenamefont {Fuhrer},
  \citenamefont {Williams},\ and\ \citenamefont
  {Ishigami}}]{chen_charged-impurity_2008}%
  \BibitemOpen
  \bibfield  {author} {\bibinfo {author} {\bibfnamefont {J.}~\bibnamefont
  {Chen}}, \bibinfo {author} {\bibfnamefont {C.}~\bibnamefont {Jang}}, \bibinfo
  {author} {\bibfnamefont {S.}~\bibnamefont {Adam}}, \bibinfo {author}
  {\bibfnamefont {M.~S.}\ \bibnamefont {Fuhrer}}, \bibinfo {author}
  {\bibfnamefont {E.~D.}\ \bibnamefont {Williams}}, \ and\ \bibinfo {author}
  {\bibfnamefont {M.}~\bibnamefont {Ishigami}},\ }\href {\doibase
  10.1038/nphys935} {\bibfield  {journal} {\bibinfo  {journal} {Nat Phys}\
  }\textbf {\bibinfo {volume} {4}},\ \bibinfo {pages} {377} (\bibinfo {year}
  {2008})}\BibitemShut {NoStop}%
\bibitem [{\citenamefont {Oostinga}\ \emph {et~al.}(2008)\citenamefont
  {Oostinga}, \citenamefont {Heersche}, \citenamefont {Liu}, \citenamefont
  {Morpurgo},\ and\ \citenamefont {Vandersypen}}]{oostinga_gate-induced_2008}%
  \BibitemOpen
  \bibfield  {author} {\bibinfo {author} {\bibfnamefont {J.~B.}\ \bibnamefont
  {Oostinga}}, \bibinfo {author} {\bibfnamefont {H.~B.}\ \bibnamefont
  {Heersche}}, \bibinfo {author} {\bibfnamefont {X.}~\bibnamefont {Liu}},
  \bibinfo {author} {\bibfnamefont {A.~F.}\ \bibnamefont {Morpurgo}}, \ and\
  \bibinfo {author} {\bibfnamefont {L.~M.~K.}\ \bibnamefont {Vandersypen}},\
  }\href {\doibase 10.1038/nmat2082} {\bibfield  {journal} {\bibinfo  {journal}
  {Nat Mater}\ }\textbf {\bibinfo {volume} {7}},\ \bibinfo {pages} {151}
  (\bibinfo {year} {2008})}\BibitemShut {NoStop}%
\bibitem [{\citenamefont {Zhang}\ \emph {et~al.}(2009)\citenamefont {Zhang},
  \citenamefont {Tang}, \citenamefont {Girit}, \citenamefont {Hao},
  \citenamefont {Martin}, \citenamefont {Zettl}, \citenamefont {Crommie},
  \citenamefont {Shen},\ and\ \citenamefont {Wang}}]{zhang_direct_2009}%
  \BibitemOpen
  \bibfield  {author} {\bibinfo {author} {\bibfnamefont {Y.}~\bibnamefont
  {Zhang}}, \bibinfo {author} {\bibfnamefont {T.}~\bibnamefont {Tang}},
  \bibinfo {author} {\bibfnamefont {C.}~\bibnamefont {Girit}}, \bibinfo
  {author} {\bibfnamefont {Z.}~\bibnamefont {Hao}}, \bibinfo {author}
  {\bibfnamefont {M.~C.}\ \bibnamefont {Martin}}, \bibinfo {author}
  {\bibfnamefont {A.}~\bibnamefont {Zettl}}, \bibinfo {author} {\bibfnamefont
  {M.~F.}\ \bibnamefont {Crommie}}, \bibinfo {author} {\bibfnamefont {Y.~R.}\
  \bibnamefont {Shen}}, \ and\ \bibinfo {author} {\bibfnamefont
  {F.}~\bibnamefont {Wang}},\ }\href {\doibase 10.1038/nature08105} {\bibfield
  {journal} {\bibinfo  {journal} {Nature}\ }\textbf {\bibinfo {volume} {459}},\
  \bibinfo {pages} {820} (\bibinfo {year} {2009})}\BibitemShut {NoStop}%
\bibitem [{\citenamefont {Feldman}\ \emph {et~al.}(2009)\citenamefont
  {Feldman}, \citenamefont {Martin},\ and\ \citenamefont
  {Yacoby}}]{feldman_broken-symmetry_2009}%
  \BibitemOpen
  \bibfield  {author} {\bibinfo {author} {\bibfnamefont {B.~E.}\ \bibnamefont
  {Feldman}}, \bibinfo {author} {\bibfnamefont {J.}~\bibnamefont {Martin}}, \
  and\ \bibinfo {author} {\bibfnamefont {A.}~\bibnamefont {Yacoby}},\ }\href
  {\doibase 10.1038/nphys1406} {\bibfield  {journal} {\bibinfo  {journal} {Nat
  Phys}\ }\textbf {\bibinfo {volume} {5}},\ \bibinfo {pages} {889} (\bibinfo
  {year} {2009})}\BibitemShut {NoStop}%
\bibitem [{\citenamefont {Mayorov}\ \emph {et~al.}(2011)\citenamefont
  {Mayorov}, \citenamefont {Elias}, \citenamefont {{Mucha-Kruczynski}},
  \citenamefont {Gorbachev}, \citenamefont {Tudorovskiy}, \citenamefont
  {Zhukov}, \citenamefont {Morozov}, \citenamefont {Katsnelson}, \citenamefont
  {Fal�ko}, \citenamefont {Geim},\ and\ \citenamefont
  {Novoselov}}]{mayorov_interaction-driven_2011}%
  \BibitemOpen
  \bibfield  {author} {\bibinfo {author} {\bibfnamefont {A.~S.}\ \bibnamefont
  {Mayorov}}, \bibinfo {author} {\bibfnamefont {D.~C.}\ \bibnamefont {Elias}},
  \bibinfo {author} {\bibfnamefont {M.}~\bibnamefont {{Mucha-Kruczynski}}},
  \bibinfo {author} {\bibfnamefont {R.~V.}\ \bibnamefont {Gorbachev}}, \bibinfo
  {author} {\bibfnamefont {T.}~\bibnamefont {Tudorovskiy}}, \bibinfo {author}
  {\bibfnamefont {A.}~\bibnamefont {Zhukov}}, \bibinfo {author} {\bibfnamefont
  {S.~V.}\ \bibnamefont {Morozov}}, \bibinfo {author} {\bibfnamefont {M.~I.}\
  \bibnamefont {Katsnelson}}, \bibinfo {author} {\bibfnamefont {V.~I.}\
  \bibnamefont {Fal�ko}}, \bibinfo {author} {\bibfnamefont {A.~K.}\
  \bibnamefont {Geim}}, \ and\ \bibinfo {author} {\bibfnamefont {K.~S.}\
  \bibnamefont {Novoselov}},\ }\href {\doibase 10.1126/science.1208683}
  {\bibfield  {journal} {\bibinfo  {journal} {Science}\ }\textbf {\bibinfo
  {volume} {333}},\ \bibinfo {pages} {860 } (\bibinfo {year}
  {2011})}\BibitemShut {NoStop}%
\bibitem [{\citenamefont {Jalilian}\ \emph {et~al.}(2011)\citenamefont
  {Jalilian}, \citenamefont {Jauregui}, \citenamefont {Lopez}, \citenamefont
  {Tian}, \citenamefont {Roecker}, \citenamefont {Yazdanpanah}, \citenamefont
  {Cohn}, \citenamefont {Jovanovic},\ and\ \citenamefont
  {Chen}}]{jalilian_scanning_2011}%
  \BibitemOpen
  \bibfield  {author} {\bibinfo {author} {\bibfnamefont {R.}~\bibnamefont
  {Jalilian}}, \bibinfo {author} {\bibfnamefont {L.~A.}\ \bibnamefont
  {Jauregui}}, \bibinfo {author} {\bibfnamefont {G.}~\bibnamefont {Lopez}},
  \bibinfo {author} {\bibfnamefont {J.}~\bibnamefont {Tian}}, \bibinfo {author}
  {\bibfnamefont {C.}~\bibnamefont {Roecker}}, \bibinfo {author} {\bibfnamefont
  {M.~M.}\ \bibnamefont {Yazdanpanah}}, \bibinfo {author} {\bibfnamefont
  {R.~W.}\ \bibnamefont {Cohn}}, \bibinfo {author} {\bibfnamefont
  {I.}~\bibnamefont {Jovanovic}}, \ and\ \bibinfo {author} {\bibfnamefont
  {Y.~P.}\ \bibnamefont {Chen}},\ }\href {\doibase
  10.1088/0957-4484/22/29/295705} {\bibfield  {journal} {\bibinfo  {journal}
  {Nanotechnology}\ }\textbf {\bibinfo {volume} {22}},\ \bibinfo {pages}
  {295705} (\bibinfo {year} {2011})}\BibitemShut {NoStop}%
\bibitem [{\citenamefont {{Castellanos-Gomez}}\ \emph
  {et~al.}(2010)\citenamefont {{Castellanos-Gomez}}, \citenamefont {Agra�t},\
  and\ \citenamefont {{Rubio-Bollinger}}}]{castellanos-gomez_carbon_2010}%
  \BibitemOpen
  \bibfield  {author} {\bibinfo {author} {\bibfnamefont {A.}~\bibnamefont
  {{Castellanos-Gomez}}}, \bibinfo {author} {\bibfnamefont {N.}~\bibnamefont
  {Agra�t}}, \ and\ \bibinfo {author} {\bibfnamefont {G.}~\bibnamefont
  {{Rubio-Bollinger}}},\ }\href {\doibase 10.1088/0957-4484/21/14/145702}
  {\bibfield  {journal} {\bibinfo  {journal} {Nanotechnology}\ }\textbf
  {\bibinfo {volume} {21}},\ \bibinfo {pages} {145702} (\bibinfo {year}
  {2010})}\BibitemShut {NoStop}%
\end{thebibliography}%

\end{document}